\documentclass[12pt]{article}
\usepackage{aasms4}

\begin{document}

\title{
Dark matter halos and the anisotropy of ultra-high energy cosmic rays
}

\author{G. A. Medina Tanco$^{1,2}$}

\author{A. A. Watson$^{2}$}

\affil{ 1.
Instituto Astron\^omico e Geof\'{\i}sico, University of S\~ao Paulo, Brasil
\\
gustavo@iagusp.usp.br
}

\affil{ 2. Dept. of Physics and Astronomy, University of Leeds,
Leeds LS2 9JT, UK \\
a.a.watson@leeds.ac.uk
}

\singlespace

\begin{abstract}

Several explanations for the existence of Ultra High Energy Cosmic
Rays invoke the idea that they originate from the decay of massive
particles created in the reheating following inflation.
It has been suggested that the decay products can explain
the observed isotropic flux of cosmic rays.  We have calculated
the anisotropy expected for various models of the dark matter
distribution and find that at present data are too sparse above
$4 \times 10^{19}$ eV to discriminate between different models.
However we show that with data from three years of operation of
the southern section of the Pierre Auger Observatory significant
progress in testing the proposals will be made.


\end{abstract}

\clearpage

\section{Introduction}

The problem of the origin of ultra high-energy cosmic rays (UHECR) is
receiving considerable attention.  The situation is very well known
and need only be summarized briefly.   Shortly after the discovery of
the cosmic background radiation Greisen (1966) and Zatsepin and Kuzmin
(1966) pointed out that interactions of cosmic ray protons and nuclei
with the 2.7 K radiation field would severely deplete the number of
events at energies beyond about $4 \times 10^{19}$ eV.  General
acceptance that events exist beyond what has come to be known as the
GZK cut-off has been long in coming but recently a consensus has
emerged that there is indeed an excess of events beyond $10^{20}$ eV
which cannot be explained by observational errors or uncertainties
in energy estimates.

Very recently (Takeda et al. 1998) the Japanese AGASA project
has reported 6 events above this energy with a spectrum which
appears to be in contradiction with what would be expected
if the sources of these particles were universal, although,
as demonstrated by Medina Tanco (1998), the number of events
is not large enough to rule out an association with nearby
extragalactic luminous matter.

The agreement of the AGASA spectrum with those from the other giant
shower detectors serves to underline the reality of the events of
greater than $10^{20}$ eV reported from them.  We note that $13$
events have been reported overall for which the energies are claimed
to be above $10^{20}$ eV: AGASA (7) (Takeda et al. 1999),
Volcano Ranch (1) (Linsley 1963), Haverah Park (4) (Lawrence,
Reid and Watson 1991), Fly's Eye (1)(Bird et al 1993) and
Yakutsk (1) (Efimov et al. 1991)).  The distribution of
events recorded by each experiment is in
reasonable agreement with their individual exposures (Watson
1998).  Not only are the particles above $10^{20}$ eV unexpected
in the face of the GZK cut-off but also many theorists find it
impossible to envisage electromagnetic methods of acceleration
to these energies.

The experimental situation with regard to the arrival direction distribution of UHECR 
is less clear cut than it is for the energy spectrum.  Using a data set dominated by 
Haverah Park events, Stanev et al. (1995) claimed that cosmic rays above $4 \times 10^{19}$ eV 
showed a correlation with the direction of the Super Galactic Plane: the level of 
significance was 2.5 - 2.8 sigma.  Later studies with AGASA data (Hayashida et al. 
1996) and with Fly's Eye data (Bird et al. 1998) did not support this claim.  Very 
recently the AGASA group (Takeda et al. 1999) have released details of 581 events 
above $10^{19}$ eV recorded by them.  Of these 47 are above $4 \times 10^{19}$ eV and 7 are above 
$10^{20}$ eV.  There is no evidence within this consistent data set to support an anisotropy 
associated with the Super Galactic Plane but they find some evidence of clustering on 
an angular scale of $2.5^{\circ}$: there are three doublets and one triplet, the chance occurrence 
of which is calculated as less than 1\%.  The triplet and a doublet, which becomes a 
triplet if a $10^{20}$ eV event from Haverah Park, lie close to the Super Galactic Plane.  
This work extends a similar earlier analysis by Uchihori et al. (1997) using a set of data 
containing events from several experiments.  If clustering of cosmic rays is established 
in very much larger data sets it will have profound implications for our ideas about 
cosmic ray origin.  For example Farrar and Biermann(1998) have claimed an 
association with radio-loud QSOs for 5 of the most energetic events.  While their 
statistical analysis has recently been challenged by Hoffman (1999), the idea is now 
capable of an independent test with the precise directions of the new AGASA events 
(Takeda et al. 1999).  So far evidence for departures from isotropy have proved 
elusive.

At  $4 \times 10^{19}$ eV about 50\% of the events are expected
to come from within $130$ Mpc while at $10^{20}$ eV the 50\%
distance is only $19$ Mpc (Hillas, 1998b).  The isotropy of
these events which must originate so close to our galaxy has
prompted a number of authors to propose that the particles may
come from the decay of super-heavy relic particles gravitationally
bound within the galactic halo.
Such super-heavy relics are postulated as having been created in
the re-heating which may follow early Universe inflation
(Berezinsky, Kacheltiess and Vilenkin (1997), Benkali, Ellis
and Nanopoulos (1998) and Birkel and Sarkar (1998)).
That such a bold hypothesis is advocated is a measure of the
difficult situation in which observation has placed theoretical
expectation.  The situation is so acute that
ideas such as the acceleration of Dirac monopoles by the galactic
magnetic field (Kephart and Weiler 1996) and the breakdown of
Lorentz invariance (Gonzalez-Mestres 1997, Coleman and Glashow
1998) are amongst those proposed to solve the enigma.

The question of super-heavy relics residing in the galactic halo and
providing a small fraction of the cold dark matter has attracted
recent attention (Berezinsky, Blasi and Vilenkin 1998,
Dubovsky and Tinyakov 1998, Hillas 1998a, Berezinsky and Mikhailov 1998
and Benson, Smialkowski and Wolfendale 1998).  In the latter two papers
estimates of the anisotropy expected have been made and Benson et al.
have compared their predictions with observation.  The present paper
extends these analyses and presents the results of the calculation in
a way which demonstrates acutely the need to have improved measurements
of the UHECR from both the Northern and the Southern Hemispheres to help
resolve the issue of a halo contribution to the UHECR.

\section{Calculations and Discussion}

\subsection{Anisotropy associated with the halo}

In what follows, we will limit the analysis to the anisotropy observed
at Earth due to the possible origin of UHECR from the decay of primaries
resident in the galactic halo. While we have been motivated by the idea
of the decay of super-heavy relic particles our results are of relevance
to any type of source of UHECR distributed throughout the galactic halo.

If UHECR are gamma-rays or neutrons, then their propagation is rectilinear
and no further assumptions are required. If, on the other hand, UHECR are
mainly charged particles, as it seems more likely
from the muon content of the largest AGASA event (Hayashida et al.1996)
and the profile of the largest Fly's Eye event (Bird et al. 1993), then they
will be deflected by the magnetic field inside the halo. In the latter case,
a good description of the topology and intensity of the halo magnetic
field, $B_{H}$, is necessary for a rigorous estimate of the anisotropy
observed at Earth. Unfortunately, there are large uncertainties regarding
$B_{H}$ (Kronberg 1994, Beck et al. 1996, Vall\'ee 1997).
However, the higher
the particle energy, the smaller the deflection.
Using and axisymmetric, spiral field without reversals and with
even (quadrupole type) parity in the perpendicular
direction to the  galactic plane (Stanev 1997),
which is consistent with the observations of our
own and other spiral galaxies (Beck et al. 1996, Kronberg 1994),
it has been shown (Medina Tanco
1997, 1998, Medina Tanco et al. 1998) that, upon traversing a $20$
kpc halo:
(a) protons with $E \sim 4 \times 10^{19}$ eV are deflected through
angles $\alpha < 10^{o}$ ($\alpha <5^{o}$ at galactic latitude
$|b| > 60^{o}$) unless their trajectories cross the central
regions of the galaxy; (b) the deflections suffered by protons
are reduced to $\alpha < 5^{o}$ at $E \sim 10^{20}$ eV for most
directions; (c) heavier nuclei, in particular Fe, are deflected by
up to $40^{o}$ for most arrival directions even at energies as high as
$E \sim 2 \times 10^{20}$ eV. In what follows only rectilinear propagation
will be considered and so, unless the UHECR are neutral, the results
should only be applied to the highest energy particles.

The emissivity of UHECR per unit volume is proportional to the number
density of potential sources in the halo, $n_{SHR}(\b{r})$ which,
in turn, we will assume to be proportional to the dark matter density
inside the galactic halo, $n_{H}(\b{r})$
where $\b{r}$ is the position vector in a galactocentric reference
system. Therefore, the incoming
flux of UHECR from a solid angle $\delta \Omega (\hat{\b{r}}')$, around
the direction $\hat{\b{r}}'$, defined in a geocentric coordinate system
is:

\begin{equation}
\delta \Phi \propto
               \int_{V_{\delta \Omega}}
               \frac{ n_{H}\left[\b{r}(\b{r}')\right]}{r'^{2}} dV
            =  \int_{0}^{ r_{H}(\b{r}')}
               n_{H}\left[\b{r}(\b{r}')\right] \delta \Omega dr'
\end{equation}

\noindent where $V_{\delta \Omega }$ is the volume of the cone of
solid angle $\delta \Omega $, $ r_{H}$ is the external radius of
the halo and $\b{r}(\b{r}')$ is the coordinate of the volume
element $dV$ in the reference system with origin on the galactic
center. Thus, the incoming UHECR flux per unit solid angle from
the direction $\b{r}'$ is:

\begin{equation}
\frac{\delta \Phi}{\delta \Omega }
                      \propto
                      \int_{0}^{r_{H}(\b{r}')}
                      n_{H}\left( 
                                 {\b{R}_{\odot}} + \b{r}'
                           \right) dr'
\end{equation}

\noindent where $\b{R}_{\odot}$ is the position of the Sun in 
the galactocentric reference system. To ensure that each direction 
on the celestial sphere has an equal weight
and that the symmetry of the problem is preserved
in the calculation 
of the anisotropy, an equal area Schmidt projection 
(Fisher, Lewis and Embleton 1993) of the sky onto 
a plane tangent to the appropriate celestial pole is used.
The projected area is populated with pixels of equal area. 
The fluxes, $\delta \Phi / \delta \Omega$, are then 
calculated for each pixel, and modulated by the exposure of a 
typical experiment, $\Xi(\delta)$, which is a function that 
depends only on declination. 
For experiments in the Northern hemisphere, the Haverah Park exposure 
at $E > 10^{19}$ eV, was used as typical, since it is located at 
latitude $54^{o}$ N, mid-way between those of AGASA ($36^{o}$ N) and 
Yakutsk ($62^{o}$ N). However Haverah Park used water-Cerenkov 
detectors so that the declination response was broader than for
the scintillator array of AGASA and Yakutsk.

The distribution of dark matter inside the halo is by no means certain. 
Nevertheless, the flatness of the rotation curves of spiral galaxies 
implies that the density inside the halo must decrease roughly as 
$1/r^{2}$. Caldwell and Ostriker (1981) parametrised the density 
of dark matter in the plane of the galaxy by a core-halo type 
model ($n_{H} \propto \left( 1+r^{2}/r_{c}^{2} \right)^{-1}$), 
and assumed that the halo is spherical (see also, Binney and 
Tremaine 1987, Sciama 1993). However, N-body simulations of 
the dissipationless formation of halos (Frenk et al. 1988, 
Katz 1991, Katz and Gunn 1991, Dubinski and Carlberg 1991, 
Dubinski 1992, Warren et al. 1992) indicate that the final 
shape is flattened. For the flattest halos obtained in the absence 
of dissipation the axial ratio, $q$, equals $0.4$. In an observational 
study of our own galaxy, van der Marel (1991) found $q > 0.34$. 

For our calculation we have assumed a bi-axial ellipsoid as an 
approximation to a flattened halo density profile; in cylindrical 
galactocentric coordinates $(\rho,\phi,z)$:

\begin{equation}
n_{H} \propto 
               \frac
               {1}
               { \left[ 
                     1+  \frac{1}{r_{c}^{2}} 
                         \left(  \rho^{2} + \frac{z^{2}}{q^{2}} \right)
     \right]  }
\end{equation}

\noindent where $r_{c}$ is a characteristic, essentially unknown, scale.
The spherical limit, $q=1$, corresponds to the isothermal halo model of
Caldwell and Ostriker (1981).

Navarro, Frenk and White (1996) (NFW), on the other hand,
investigated the structure of dark halos in the standard Cold
Dark Matter model, and
found that the spherically averaged density profile can be fit over
an interval of two decades in radius by scaling a "universal" profile.
Their halo profiles are approximately isothermal over a large range
in radii, but shallower than $r^{-2}$ in the central region and
steeper than $r^{-2}$ near the virial radius:

\begin{equation}
n_{H} \propto \frac{1}
                   {
                    \frac{r}{r_{s}} \left(  1+ \frac{r}{r_{s}}  \right)^{2}
                   }
\end{equation}

\noindent where $r_{s}$ is a characteristic radius (not the halo core).
In our analysis we consider halo profiles given by both eq. (3) and
(4).

Figure 1 shows a step-by-step graphical description of our procedures.
In figure 1a the density profile, given by (3) with $q=0.4$ and
$r_{c}=8$ kpc is shown. The horizontal axis, $\rho$, runs along
the galactic plane, while the vertical axis, $z$, is perpendicular
to the galactic plane. Figure 1b shows the flux of UHECR per unit
solid angle, originated by the density profile in 1a, in galactic
coordinates with the galactic centre at the
center of the figure. Figure 1c shows  $\delta \Phi / \delta \Omega$
from figure 1b rotated into equatorial coordinates. Figures 1d and
1f are the Schmidt projections of $\delta \Phi / \delta \Omega$ from
1c onto planes tangent to the North and South pole respectively.
Figures 1e and 1g show the Schmidt projections 1d and 1f convoluted
with the response in declination of Haverah Park ($54^{o}$ N) and
Auger South (Malarg\"ue, Argentina) respectively.
For the Malarg\"ue site ($35^{o}$ S) we have used the Haverah Park
declination distribution (appropriately mirrored and shifted) as the
actual declination distribution has yet to be measured and
the Pierre Auger Observatory will use water-Cerenkov tanks of the
same depth as those used at Haverah Park.
It is from these later figures, and similar ones for other halo models,
that the anisotropies discussed below has been calculated.

We have used the amplitude and phase of the first harmonic to
characterize the anisotropies. Thus (e.g., Linsley 1975),
the amplitude is:

\begin{equation}
r_{1h} = \sqrt{ a_{1h}^2  +  b_{1h}^2}
\end{equation}

\noindent where:

\begin{equation}
a_{1h} = \frac{2}{N} \sum_{i=1}^{N} cos \alpha_{i}
\; \; \; \; \mbox{,} \; \; \; \;
b_{1h} = \frac{2}{N} \sum_{i=1}^{N} sin \alpha_{i}
\end{equation}

\noindent the phase is

\begin{equation}
\Psi_{1h} =\mbox{tan}^{-1} \left( \frac {b_{1h}} {a_{1h}} \right)
\end{equation}

\noindent and $\alpha_{i}$ is the right ascension of an individual
event.

The rms spread in amplitude and phase of the first harmonic are
given by:

\begin{equation}
\Delta r = \sqrt{\frac{2}{N}}
 \end{equation}

\noindent and

\begin{equation}
\Delta \Psi = {\frac{1}{\sqrt{2 k_{0}}}}
\end{equation}

\noindent where $k_{0} = r_{1h}^{2} N / 4$.
Another quantity of interest is the number of events
required for a signal-to-noise ratio of $ n_{\sigma} $
standard deviations either in amplitude or phase:

\begin{equation}
N_{r}( n_{\sigma}) = \frac{2 n_{\sigma}^{2}}{r_{1h}^{2}}
\; \; \; \; \mbox{,} \; \; \; \;
N_{\Psi}( n_{\sigma}) = \frac{2 n_{\sigma}^{2}}{r_{1h}^{2} \Psi_{1h}^{2}}
\end{equation}

In figure 2 $ N_{r}( n_{\sigma}=3) $ is shown for the set of models
described by equation(3) as a function of the characteristic
length $r_{c}$ and different values of $q$, covering very flat
solutions ($q=0.2$) to the isothermal solution ($q=1.0$).
The magnitude of $r_{1h}$ depends on the halo model: for the
models described by equation (3) $r_{1h}$ decreases as $q$
increases at constant $r_{c}$, while at constant $q$, $r_{1h}$
decreases as $r_{c}$ increases.
The curves have been calculated for Haverah Park, but they are
also representative of what would be expected for
AGASA and Yakutsk. We note that the grand
total number of events with $E > 4 \times 10^{19}$ eV for the
Northern Hemisphere sites is $N \sim 100$. Therefore, it is not
possible, with the present data
to measure the amplitude of the first harmonic at the
$3 \sigma$ level required to have statistically significant
discriminators between any dark halo model density profiles.

Figure 3 shows phase vs. amplitude of the first harmonic for dark
halo models (3) and (4) (NFW) for $2 < r_{c} < 50$ kpc and
$10 < r_{s} < 100$ kpc respectively. For model (3) flattenings
$0.2 \le q \le 1$ are shown. For every model, the larger the
amplitude of the first harmonic the more centrally concentrated
is the halo (i.e., smaller $r_{c}$ or $r_{s}$). The error bars represent
68\% confidence levels for Volcano Ranch (6 events, Linsley 1980)
Haverah Park (27 events, Reid and Watson 1980), Yakutsk
(24 events, Afanasiev et al. 1995) and AGASA
(47 events, Hayashida et al. 1996, Uchihori et al. 1997,
Takeda et al. 1999)
at $E > 4 \times 10^{19}$
eV, and 95\% confidence for the 104 events of the four experiments
combined. For the latter the error box is also shown in shades
of gray in the background. Note the strong increase of the uncertainty
range in phase as the amplitude decreases.
It is evident that the data available at present are
insufficient to restrict any particular dark matter halo model.
At most it can be said that the data are not incompatible with
UHECR originating in a spherical, or only slightly flattened halo
($q > 0.6$).
An isothermal halo is as acceptable as, and is indistinguishable
from, a NFW type of halo model, regardless of the value of their
characteristic scales. Furthermore, the number of events detected
so far by each experiment is so small that statistical fluctuations
may even dominate the results.

Figures 4 and 5 show how much the situation can improve using the
Southern site of the Auger experiment (Malarg\"ue, Argentina,
$\sim 35^{o}$
South) which is to be developed. Comparing figures 3 and 5 it is
evident that an experiment
located in the Southern Hemisphere has a larger potential to
discriminate between halo models than one located in the Northern
hemisphere for small $N$, provided
$r_{c} {\stackrel{\sim}{>}} 10$ kpc. Location
is not enough, however,
and figures 2 and 4 imply that a significantly larger exposure is
needed to make a difference from the current status. After three
years of operation of the $3000$ km$^{2}$ Southern hemisphere
Auger detector, roughly $\sim 570$ events are expected above
$4 \times 10^{19}$ eV, and that should allow $3 \sigma$ amplitude
determinations for the flatter halo models
(the constraints on phase are always smaller).
As an example, suppose that a measured harmonic amplitude
is regarded as being established when the probability that it
could have arisen from a random distribution through a chance
fluctuation is less than $10^{-3}$.
It follows that with 500 events an amplitude of $24$\% would
be detectable and the phase would have an uncertainly of
$\pm 15^{o}$.
Simulated error boxes are shown in figure 5
for this supposed amplitude and for one of $70$\%.
It is clear from the figure that such a
result would eliminate a number of halo possibilities depending
on the value of the phase which is measured.
Therefore, after three years of operation,
it should be possible to exclude some dark halo
models.

\subsection{Anisotropy associated with Andromeda (M31)}

It is a well known fact in gamma ray burst
research, that a halo origin of the bursts is ruled out by the
non-observation of clustering of events in the direction of
Andromeda galaxy (M31, the largest galaxy in the local group
at a distance of only $D \sim 670$ kpc). That much the same
reasoning should apply to the present UHECR problem has been
most recently discussed by several authors (Benson, Smialkowski and
Wolfendale 1998, Dubovsky and Tinyakov 1998). However, very
different values have been quoted in these works for the
contribution of Andromeda in UHECR.
We have therefore made an independent calculation of the magnitude
of the effect.
The ratio between the
incoming UHECR flux originating in Andromeda and that originating
in the halo of our own galaxy inside a given solid angle
$\delta \Omega $ can be expressed as:

\begin{equation}
\frac{\Phi_{M31}}{\Phi_{MW}} \sim  \frac{\zeta}{D^{2}} \times
           \frac{\int_{V_{H}} n_{H} dV }
           {\int_{V_{\delta \Omega}} \frac{n_{H}}{r^{2}} dV }
 \end{equation}

\noindent where the second factor
on the right hand side of the equation
is a function that depends
only on the particular halo model assumed and
$\zeta \sim 2$ is the ratio between the masses of the halos of
Andromeda and the Milky Way. The integration volume $V_{H}$ is
the volume of the Galaxy halo and $V_{\delta \Omega}$ is
the volume defined by the cone of solid angle $\delta \Omega$
pointing in the direction of Andromeda.

Figure 6 shows $\Phi_{M31} / \Phi_{MW}$ for a
$10^{o} \times 10^{o}$ solid angle (the expected spread due to
deflection of a $4 \times 10^{19}$ eV proton arriving from
Andromeda - e.g., Medina Tanco et al. 1997) for several
isothermal (i.e., $q=1$ in eq. (3)) halo models. The three
models have been normalized in such a way as to give the
same contribution to the galactic rotation curve at a
galactocentric distance of $r_{o} = 18$ kpc and differ in the
ratio $\eta$ between the total halo mass and the dark matter
mass inside $r_{o}$. Galactic dark halos with $\eta = 2$, $5$
and $10$ are considered.
The results show that the contribution from Andromeda increseas
faster than the contribution from our own galaxy as the mass
of the halo is increased.
Due to the limited size of the present UHECR sample
($\sim 0.5$ events per  $10^{o} \times 10^{o}$ solid angle),
nothing can yet be said about the
existence of an UHECR contribution originated in the dark
halo of Andromeda.

\section{Comments on related work}

Other authors have recently discussed the anisotropy expected if the
UHECR are produced by the decay of super-heavy relic particles in the
galactic halo (or indeed by any other sources distributed in a similar
way).  Berezinsky and Mikhailov (1998) have used the Isothermal
distribution
of dark matter (Kravtsov et al 1997) and the distribution predicted by
the numerical simulations of Navarro, Frenk and White (1996) to predict
the amplitudes of the first and second harmonics and the phase of the
first harmonic for the geographical location of the Yakutsk array
(latitude = $62^{o}$ N).  This is an extension of the calculation
outlined in Berezinsky, Blasi and Vilenkin (1998) in which a wide-ranging
overview of the signatures from topological defects is given. The
amplitudes and phases which they predict are very similar to those
found in our calculation (figure 3).  For the Isothermal model they
calculate the phase to be $250^{o}$ and find that the amplitude of the
first harmonic varies from $0.40$ to $0.14$ as $r_{c}$ changes from
5 to 50 kpc.  For the NFW model the same phase is found and the harmonic
amplitude varies form $0.38$ to $0.31$ as $r_{s}$ changes from $30$ to
$100$ kpc. These results are in reasonable agreement with our
work (figure 3).

Berezinsky and Mikhailov state that dominance of a halo component at
about $10^{19}$ eV can probably be excluded by the AGASA data which
contains nearly $600$ events above this energy.  However in our view
this is not a very strong conclusion as there is no acute problem at
$10^{19}$ eV comparable to that which exists at higher energy.
Particles of $10^{19}$ eV can probably be produced in several locations
by electromagnetic processes. Additionally there is no difficulty in explaining the isotropy as a reasonable fraction of 
the particles may be iron nuclei.  This is allowed by the necessarily model-interpretation of the Fly's Eye data and the limited statistics (Bird et al. 1995, Ding et 
al., 1997).  Iron nuclei cannot, of course, be created by the decay of dark matter 
particles.

Benson, Smialkowski and Wolfendale (1998) have used data from a variety
of experiments to discuss the dark matter contribution from two halo
possibilities, one in which an extensive ($100$ kpc) magnetic halo is
postulated and one in which the dark matter density distribution follows
the Navarro, Frenk and White (1996) model.   They make comparisons with
their predictions using data  at $(1 - 5) \times 10^{18}$ eV from Akeno
and Yakutsk and above $3 \times 10^{19}$ eV using data from AGASA, Volcano
Ranch, Haverah Park, Yakutsk and Sydney as discussed in Chi et al. (1992).

It does not seem possible, to us, to extract meaningful information on
the super-heavy relic content of the halo from the arrival direction
distribution of events as low in energy as $(1 -5) \times 10^{18}$ eV.
Here there are likely to be many iron nuclei present and, as at
$10^{19}$ eV, there is no enigma to be resolved which
necessitates the postulate of dark matter particles.

In our discussion of the data above $4 \times 10^{19}$ eV we have
used the $104$ events (figure 3) from Volcano Ranch, Yakutsk,
Haverah Park and AGASA.  We have shown that this
number of events is insufficient to discriminate against models
other than those with rather flat distributions ($q<0.4$).
We believe that it is inappropriate to try to draw conclusions using
observations made with the Sydney array as Benson, Smialkowski and
Wolfendale(1998) have attempted.  Of the 80 events with energies
above $4 \times 10^{19}$  eV in the Sydney catalogue, $60$ have
zenith angles smaller than $60$ degrees.  However the mean
multiplicity of struck stations in only $5.0$ and $60$\% of these
events have $3$ or $4$ fold multiplicity.  This means that the
core location, and hence the reconstructed muon size, is very
uncertain.  There are also well-documented difficulties with the
instrumentation of the Sydney experiment (e.g. Watson 1991) and
with the models used to estimate the energies (Hillas 1990).
The conclusions reached by Chi et al. (1992) about the Sydney
data result in an energy spectrum (figure 7a of Chi et al.) which
is not consistent with the modern spectra from
AGASA, Haverah Park and Fly's Eye. For several reasons, therefore,
we deem it prudent to ignore those data.

\section{Conclusions}

We have calculated the anisotropy of UHECR  to be expected at
specimen locations in the Northern and Southern hemispheres on
the assumption that the particles are created in the decay of
super-heavy relic particles within the galactic halo.  Several
models describing the distribution of  cold dark matter have been
considered.  We conclude that our calculations are in good agreement
with other work but that it is premature to draw inferences about the
existence, or otherwise, of sources of UHECR lying within the halo of
our galaxy.  The issue could be resolved relatively quickly by the
Pierre Auger Observatory, construction of the Southern part of which
is scheduled to begin in 1999.

\bigskip

\section{Acknowledgments}

GAMT is also grateful to the High-Energy Astrophysics group of the
University of Leeds for its kind hospitality.
This work was partially supported by the Brazilian agency FAPESP
through a fellowship to GAMT.



\newpage

\noindent
{\bf Figure Captions}

\bigskip

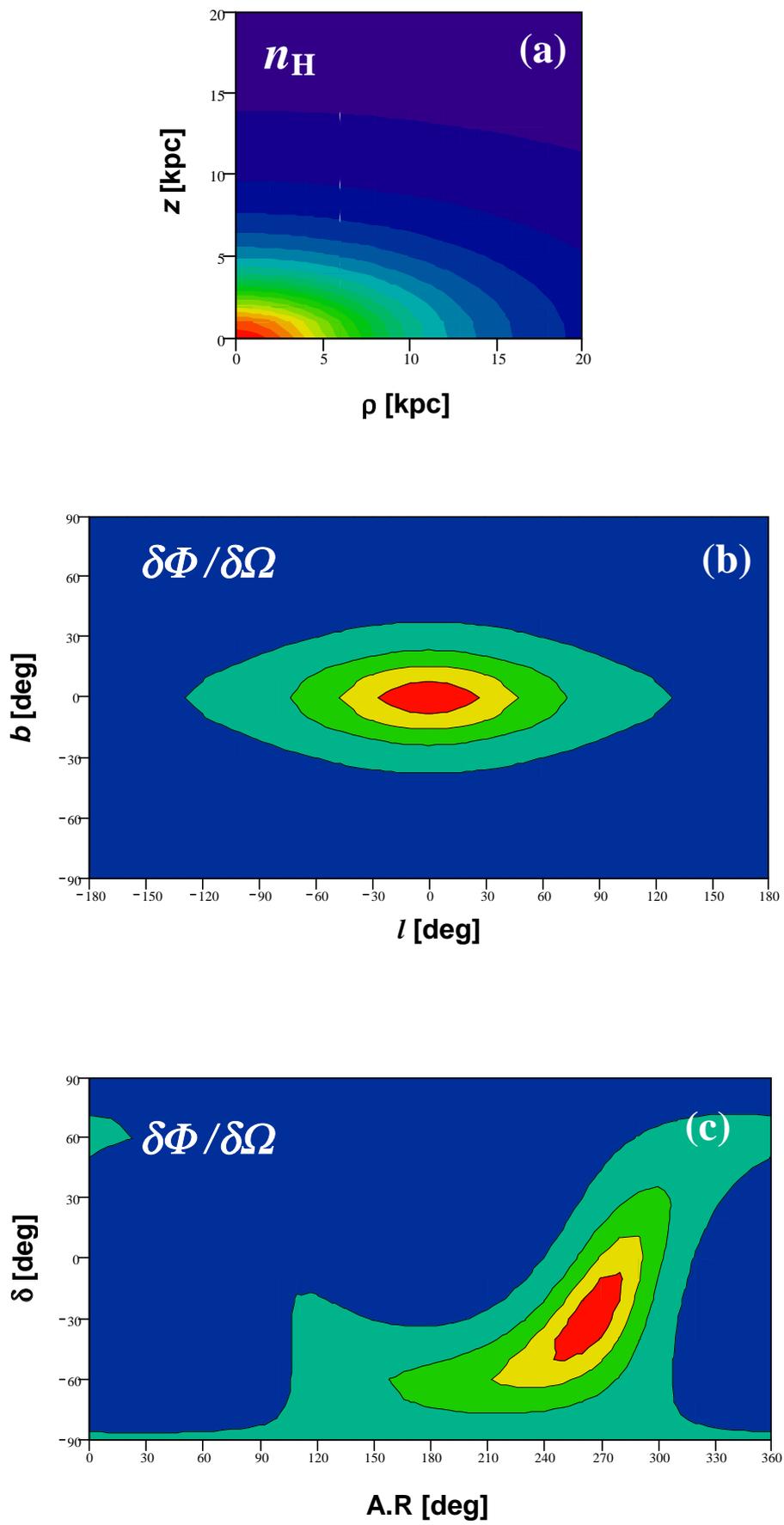
{\bf Figure 1:} : A graphical example of the procedure followed is shown.
(a) Halo density (cylindrical galacto-centric coordinates) given by eq.(3)
with $q=0.4$ and $r_{c}=8$ kpc; distances are in kpc and density contours
are
linear; the density distribution is shown for one-quarter of the Galaxy.
(b) UHECR flux produced by dark matter distribution (a) as seen
in galactic coordinates. (c) As (b) but in equatorial coordinates.
(d) Schmidt projection of (c) onto the North Pole. (f) As (d) but
for the South Pole. (e) and (g) are the projections (d) and (f)
convoluted with the response in declination of Haverah Park and
Auger South respectively.
First harmonics have been calculated over figures of type (e)
and (g) for a variety of halo models.


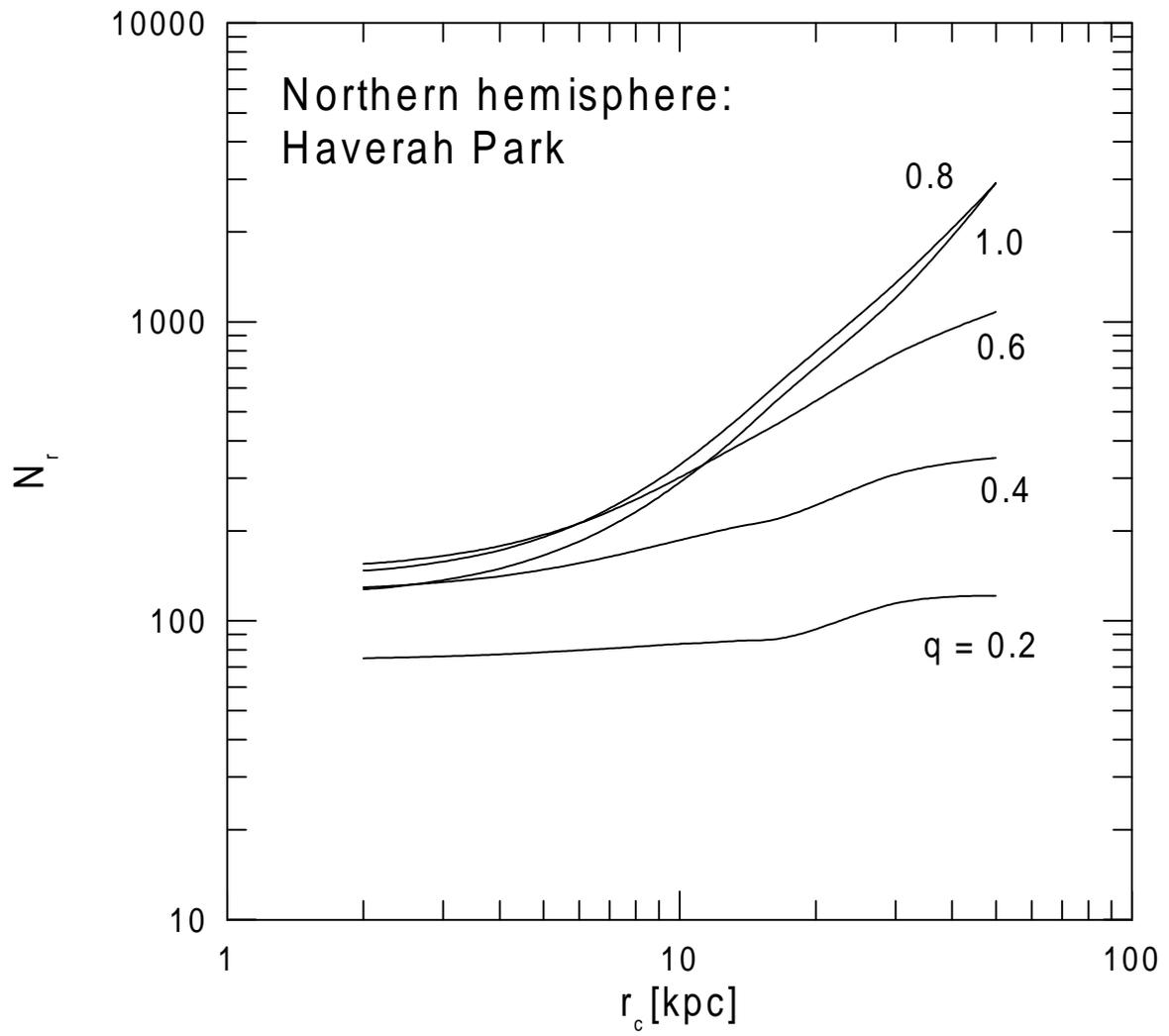
{\bf Figure 2:} Number of events necessary for an
amplitude determination significant at the $3 \sigma$ level for
several halo models. Note that the
existing Northern hemisphere database (AGASA, Haverah Park
Volcano Ranch and
Yakutsk) at $E > 4 \times 10^{19}$ eV comprises only 104 events.


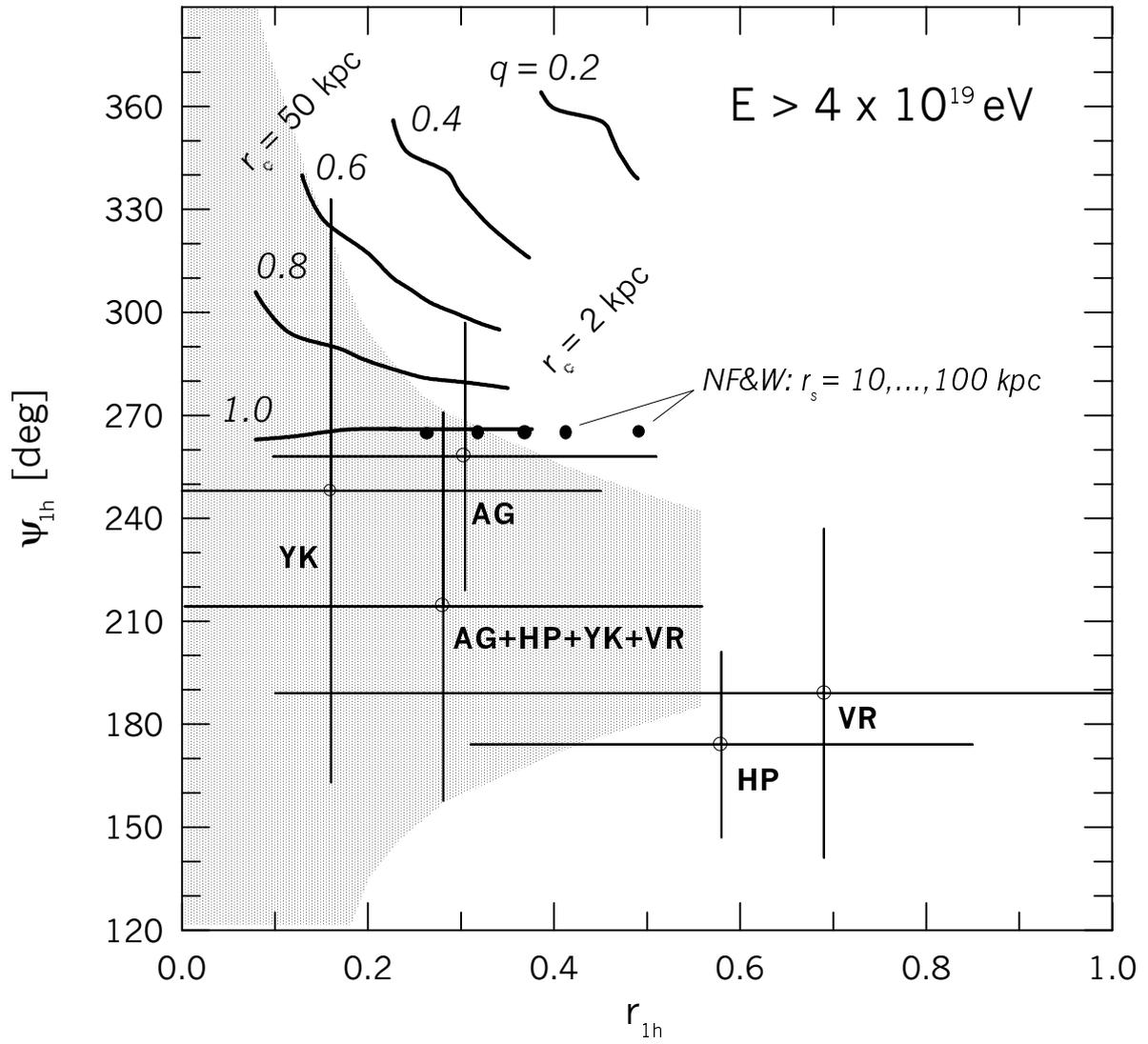
{\bf Figure 3:} Phase versus amplitude of the first harmonic for
the several models described in the text.
The heavy dots are NFW models for $r_{s} = 10$, $20$, $30$, $50$
and $100$ kpc. The lines identify models described by equation (3)
for $2 \le r_{c} \le 50$ kpc and $0.2 \le q \le 1.0$.
$r_{s}$ and $r_{c}$ are explained in the text.
Error bars correspond to
68\% C.L. for the available data from Volcano Ranch (VR, 6 events),
Yakutsk (YK, 24),
AGASA (AG, 47) and Haverah Park (HP, 27) with
$E > 4 \times 10^{19}$ eV. The 95\% C.L. error bars for the
combination of the experiments (AG+HP+YK+VR, 104) is
also shown.
The shaded region denotes the 95\% C.L. combined error box, and
stresses the increase of the error in phase as the amplitude
decreases.


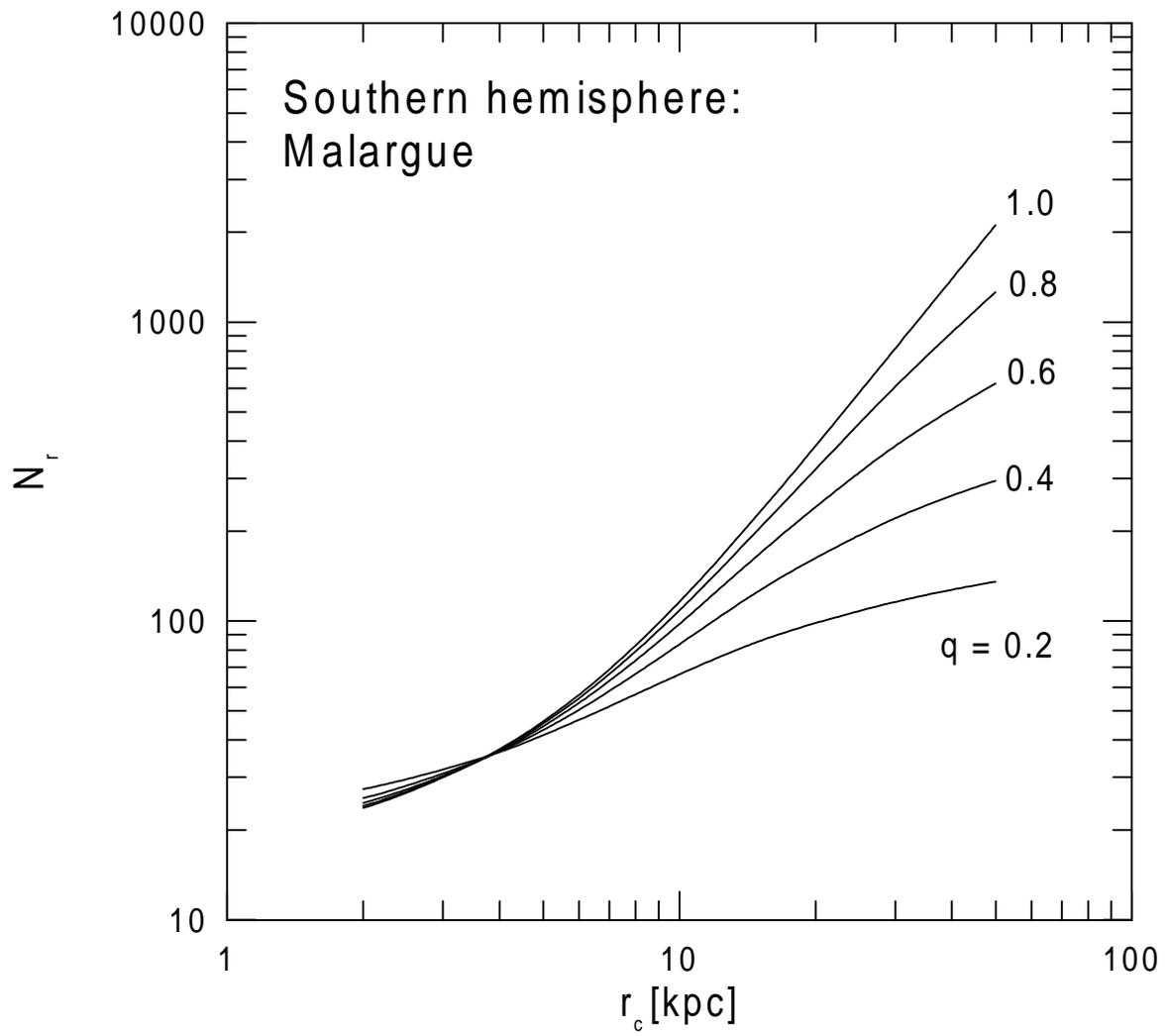
{\bf Figure 4:} Same as figure 2 but calculated for Malarg\"ue,
Argentina, the Southern site of the Auger experiment.


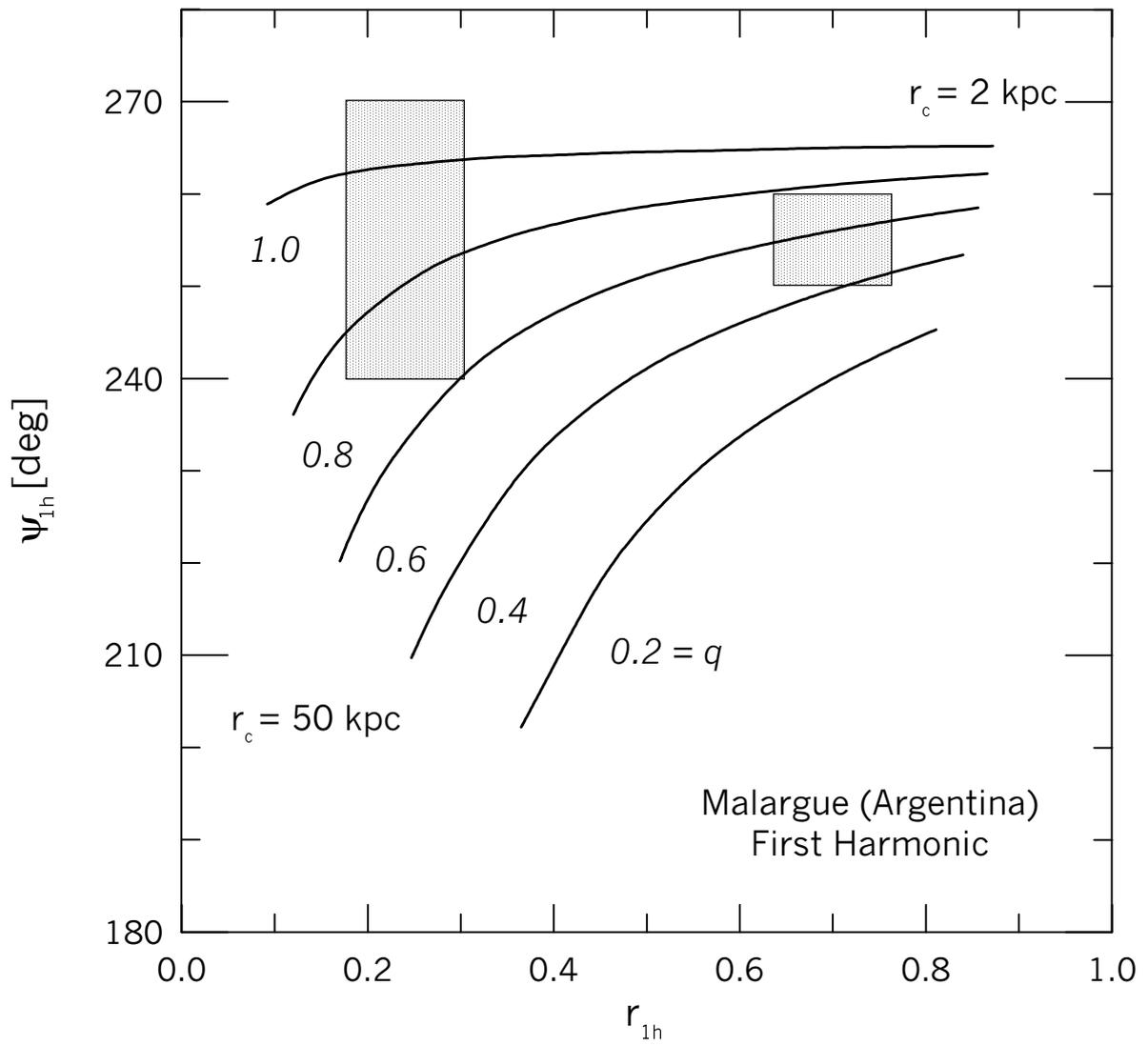
{\bf Figure 5:} Same as figure 3 but calculated for Malarg\"ue.
The error boxes are two simulated $68$\% C.L. data points,
corresponding to hypothetical first harmonic amplitudes equal to
$0.24$ and $0.7$ respectively as would be found after $3$ years
of observation
(i.e., $\sim 500$ events with $E > 4 \times 10^{19}$ eV).


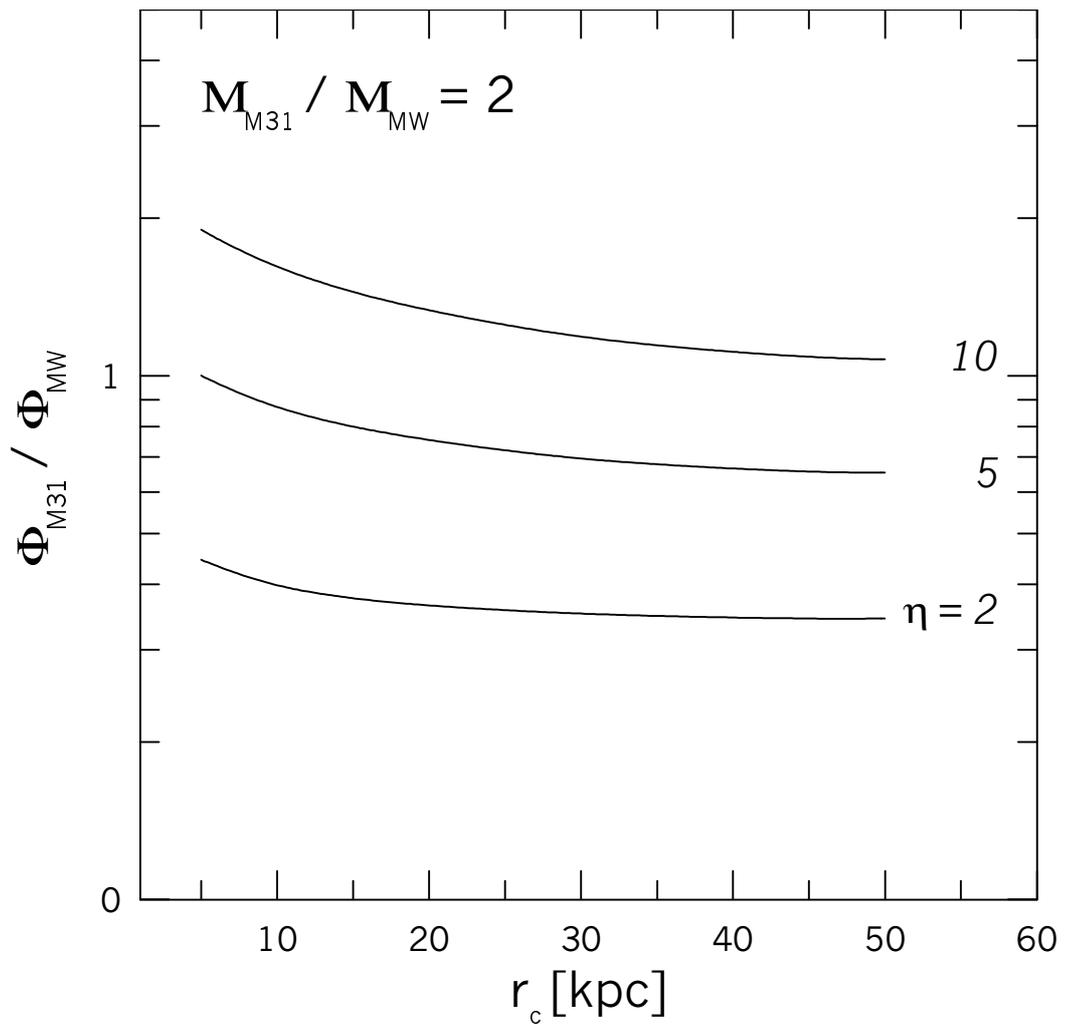
{\bf Figure 6:} The contribution of Andromeda (M31). Ratio between
the flux of UHECR originating in the halo of Andromeda and in
our own halo, within  a cone of $10^{o} \times 10^{o}$ centered
in the direction to M31.
The calculations shown are for the isothermal halo (eq. (3) with
$q=1$). The models are normalized to reproduce the galactic rotation
curve inside $r_{o} \sim 18$ kpc, but differ in the total mass
of the Galaxy halo, $M_{MW} = \eta \times M(r \le r_{o})$,
where $\eta$ is the mass of our halo in units of the mass inside
$r_{o} = 18$ kpc.
At present, the average number of UHECR
detected above $E > 4 \times 10^{19}$ eV is only $\sim 0.5$ events
on a sky area of $10^{o} \times 10^{o}$ so not conclusion may be
drawn.

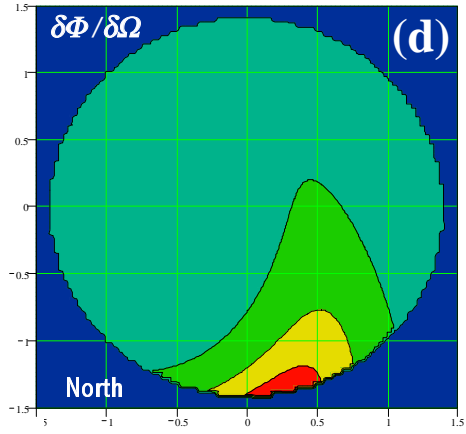
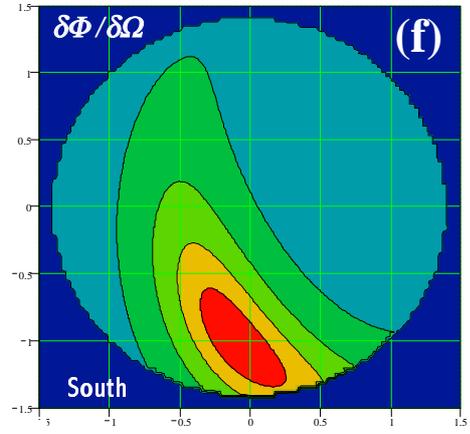
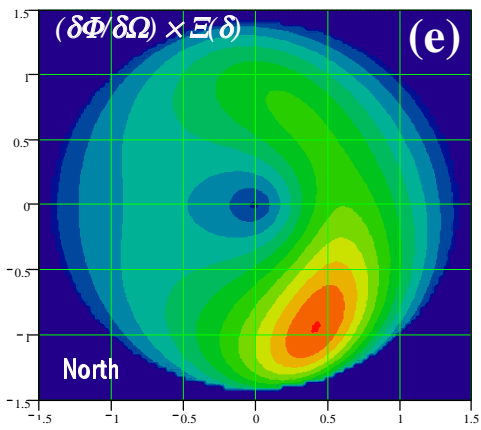
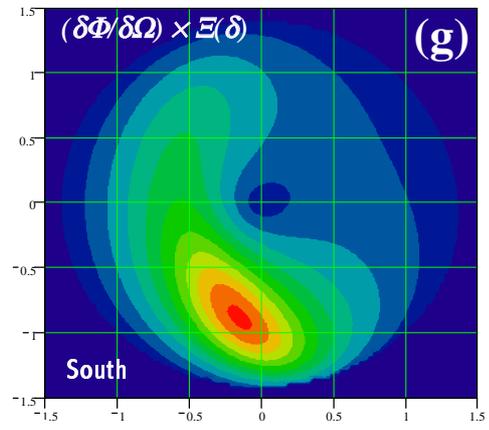
\end{document}